\documentclass[10pt]{article}
\usepackage[OE]{express}
\usepackage{graphicx}

\usepackage{bm}
\usepackage{amsmath}
\usepackage{bbold}
\usepackage[utf8]{inputenc}

\begin{document}
\title{Topological dynamics and excitations in lasers and condensates with saturable gain or loss}

\author{Simon Malzard,\authormark{1} Emiliano Cancellieri,\authormark{1,2} and Henning  Schomerus\authormark{1,*}}

\address{\authormark{1}Department of Physics, Lancaster University, Lancaster, LA1 4YB, UK\\
\authormark{2}Department of Physics and Astronomy, University of Sheffield, Sheffield, S3 7RH, UK}

\email{\authormark{*}h.schomerus@lancaster.ac.uk} 



\begin{abstract*}
We classify symmetry-protected and symmetry-breaking dynamical solutions for nonlinear saturable bosonic systems that display a non-hermitian charge-conjugation symmetry, as realized in a  series of recent groundbreaking experiments with lasers and exciton polaritons. In particular, we show that these systems support stable symmetry-protected modes that mirror the concept of zero-modes in topological quantum systems, as well as symmetry-protected power-oscillations with no counterpart in the linear case.
In analogy to topological phases in linear systems, the number and nature of symmetry-protected solutions can change. The spectral degeneracies signalling phase transitions in linear counterparts extend to bifurcations in the nonlinear context. As bifurcations relate to qualitative changes in the linear stability against changes of the initial conditions, the symmetry-protected solutions and phase transitions can also be characterized by topological excitations, which set them apart from symmetry-breaking solutions.
The stipulated symmetry appears naturally when one introduces nonlinear gain or loss into spectrally symmetric bosonic systems, as we illustrate for one-dimensional topological laser arrays with saturable gain and two-dimensional flat-band polariton condensates with density-dependent loss.
\end{abstract*}

\ocis{(140.3430)   Laser theory; (140.3290)  Laser arrays; (190.3100) Instabilities and chaos;  (270.2500) Fluctuations, relaxations, and noise.} 

%

\section{Introduction}

A wide range of topological quantum effects manifest themselves in symmetries of an excitation spectrum.
This relation has been  explored  extensively for fermionic systems, for which the single-particle Hamiltonian
can obey a chiral symmetry or a charge-conjugation symmetry \cite{RevModPhys.82.3045,RevModPhys.83.1057,beenakker_random-matrix_2015}. Allowing also for time-reversal invariance one can identify ten universality classes featuring a variety of topological quantum numbers. These quantum numbers determine the formation of zero modes and unidirectional transport channels at edges, surfaces, and interfaces of various dimensions \cite{1367-2630-12-6-065010,PhysRevB.82.115120}. In the fermionic case these symmetries are fundamental as they are owed to the structure of Fock or Nambu space \cite{Heinzner2005}, and therefore also extend to the interacting case \cite{PhysRevB.83.075103}. Prime examples are superconducting systems, for which a charge-conjugation symmetry dictates that excitations $\psi\exp(-i\omega t)$ at a positive frequency $\omega$ are paired with excitations $X \psi^*\exp(i\omega t)$ at the corresponding negative frequency $-\omega$, where $X$ is a suitable unitary transformation. The systems can then also feature robust Majorana zero modes $\psi=X\psi^*$ at $\omega=0$ \cite{0034-4885-75-7-076501,0268-1242-27-12-124003,doi:10.1146/annurev-conmatphys-030212-184337}, an effect which translates to the fermion parity anomaly in the full many-body theory \cite{1063-7869-44-10S-S29,PhysRevB.79.161408,PhysRevLett.110.017003}.

Some of these symmetries also play a natural role beyond the fermionic context.  For instance, time-reversal symmetry is intimately related to optical reciprocity
\cite{0034-4885-67-5-R03}, a classical concept whose modification leads to topological effects in photonic structures \cite{lu_topological_2014} and analogous optical \cite{nori2015,peano_topological_2016,PhysRevX.5.031011,PhysRevX.6.041026}, acoustic \cite{Suesstrunk47,PhysRevLett.114.114301} and mechanical \cite{kane_topological_2014} systems.
Topological effects can also be engineered into bosonic quantum systems, such as cold atomic gases
\cite{goldman_topological_2016} and exciton polaritons \cite{PhysRevX.5.011034,PhysRevX.5.031001,PhysRevLett.114.116401}.
For weakly interacting bosons the  excitations are again characterized by a Bogoliubov theory \cite{pitaevskii_bose-einstein_2003,RevModPhys.78.179,PSSB:PSSB200560961,Kawaguchi2012253},
and recent work has explored a variety of mechanisms to engineer topological features into this description \cite{PhysRevA.88.063631,PhysRevB.87.174427,PhysRevA.91.053621,PhysRevB.93.020502,1367-2630-17-11-115014,PhysRevLett.115.245302,PhysRevLett.117.045302}.
The charge-conjugation symmetry can also be induced into the single-particle theory of bosonic systems, where one again can admit linear gain or loss \cite{PhysRevLett.102.065703,PhysRevLett.110.013903,schomerus_topologically_2013,poli_selective_2015,arXiv:1704.00896}.
The spectrum of the  effective Hamiltonian then displays symmetric pairs of complex frequencies $(\Omega_n,-\Omega_n^*)$, which can be interpreted as resonances. Topological features persist because this spectral constraint can enforce a number of purely imaginary self-symmetric resonances $\Omega_n=-\Omega_n^*$, which provide an analogue to broadened fermionic zero modes \cite{Pikulin2012a,PhysRevB.87.235421,PhysRevLett.115.200402,sanjose2016,sanjose2018}.

Five recent experiments aimed at realizing topological zero modes in polaritonic condensates \cite{2017arXiv170407310S,2017arXiv170503006} and lasers  \cite{R1,R2,R2a}. As these are inherently nonlinear systems, the question arises whether the notion of zero modes and topological protection persists.
Here, we provide a unifying perspective on these systems based on a notion of charge-conjugation symmetry that applies directly to the time-dependent nonlinear wave equation.
Our general strategy is as follows: Topological states in linear systems are protected by symmetry, but their number can change discretely in phase transitions, which are generally linked to degeneracies (such as when a band gap closes). This notion is underpinned by the continuity of the spectrum under smooth parameter changes (deformations of the system), a feature at the heart of linear spectral analysis. Analogously, we show that nonlinear systems can display dynamical solutions that are protected by symmetry. Their number can change in dynamical degeneracies, which correspond to bifurcations. This notion is then underpinned by the structural stability of dynamical solutions, which is at the heart of nonlinear dynamics. This nonlinear notion of topological states reduces to the conventional spectral notions in the linear limit. New effects also appear: We identify a new class of topological solutions that does not have a linear counterpart, and also exploit the link of bifurcations to a qualitative change of the linear stability against changes of initial conditions, which again does not feature in the linear context.

In detail, as we will see, nonlinearities in the gain or loss  lead to complex-wave dynamics where the time-dependent solutions appear in pairs $\Psi(t)$ and $X\Psi^*(t)$. In contrast to a general time-reversal symmetry, which induces a partner solution $X\Psi^*(-t)$, the solutions $\Psi(t)$ and $X\Psi^*(t)$ both exhibit the same arrow of time.
This time-preserving symmetry therefore allows for self-symmetric states---including stationary states $\Psi(t)=\Psi_0=X\Psi_0^*$, which we interpret as zero modes. As a notable feature without analogue in the linear case, the symmetry also protects a twisted variant of time-dependent power-oscillating states with $\Psi(t+T/2)=X\Psi^*(t)$, for which periodicity $\Psi(t+T)=\Psi(t)$ is enforced by symmetry---suggesting that the nonlinear setting admits for a richer notion of protected states than the linear case. That these states enjoy topological protection can be further ascertained by identifying topological excitations in the stability spectrum, which we find to govern the spectral phase transitions between the different types of symmetry-protected solutions---corresponding to different topological phases of the dynamical system.
We illustrate our findings for  two model systems representing one-dimensional topological laser arrays with saturable gain and two-dimensional flat-band polariton condensates with density-dependent loss.

\section{Results}
\subsection{Model and classification of states}

Consider a bosonic system whose classical limit is described by a complex-wave equation
\begin{equation}
i\frac{d}{dt}\Psi(t)=H\Psi(t)+F[\Psi(t),\Psi^*(t)]\Psi(t).
\label{eq:nonlinwaveeq}
\end{equation}
The operator $H$ provides an effective single-particle description, while the functional $F$ describes the nonlinear effects which may encompass gain and loss.  We assume these effects to be local in time and often will drop the time argument.
In keeping with the quantum origin of the wave function, the nonlinear effects should preserve the global $\mathrm{U}(1)$ gauge freedom, so that any solution can be multiplied by a fixed phase factor $\exp(i\alpha)$. This can be guaranteed by assuming $F_\Psi \Psi=F_{\Psi^*} \Psi^*$, where subscripts denote functional derivatives.
In addition, we here stipulate that $H$ displays a charge-conjugation symmetry, $XHX=-H^*$ where $X$ is a unitary operator with $X^2=1$. To extend this notion to the nonlinear case, we similarly demand that
$XF[\Psi,\Psi^*]X=-(F[X\Psi^*,X\Psi])^*$, with the same operator $X$.

\begin{figure*}[t]
\includegraphics[width=\textwidth]{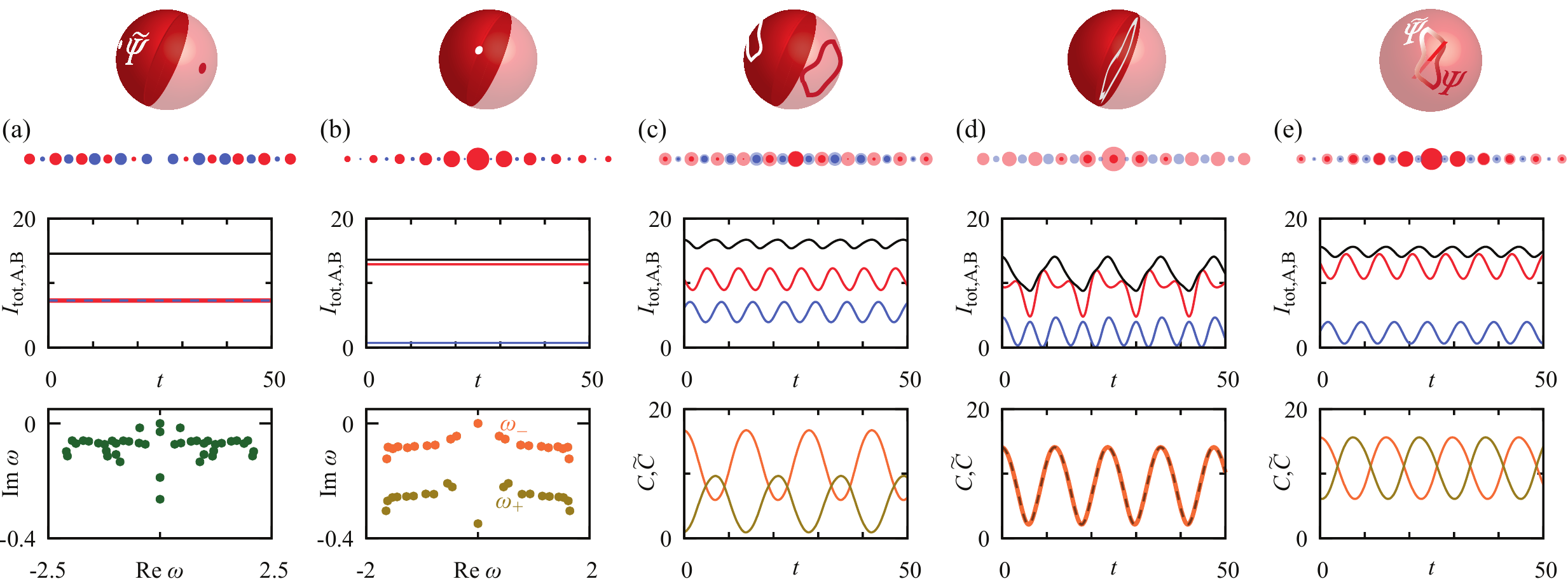}
\caption{\label{fig1}
Illustration of the five types of modes for a topological laser array based on a Su-Schrieffer-Heeger chain with background loss $\gamma_A=\gamma_B=0.3$ and various amounts of saturable gain [see Eqs.~\eqref{eq:TB1}, \eqref{eq:TB2} and \eqref{eq:saturableterms1}].  (a) Stationary symmetry-breaking mode ($g_A=0.4$, $g_B=0.7$). (b) Stationary self-symmetric zero mode ($g_A=0.8$, $g_B=0.0$). (c) Oscillating symmetry-breaking mode ($g_A=0.7$, $g_B=0.4$). (d) Oscillating self-symmetric mode ($g_A=0.7$, $g_B=0.4$ with symmetry-preserving initial conditions). (e) Twisted oscillating mode ($g_A=0.8$, $g_B=0.2$).
The sketches at the very top symbolize the traces of the solutions in low-dimensional cross-sections of the dynamical phase space, where in (a-d) the symmetry is represented as a reflection and in (e) as a rotation. In the second row, the circles represent the resonators, where the area denotes the intensity (A and B sublattice in red and blue; in the time-dependent case, we show two circles corresponding to the largest and smallest intensity over a cycle.)
The third row shows the time traces
of the intensities $I_{A}=|\mathbf{A}|^2$ (red), $I_{B}=|\mathbf{B}|^2$ (blue), and $I_{\rm tot}=I_A+I_B$ (black).
The bottom panels in (a,b) show the stability spectra of the stationary states, while in (c-e) they show
the correlation functions $C=|\langle \Psi(0)|\Psi(t)\rangle|$ (orange),
$\tilde C=|\langle \Psi(0)|\tilde \Psi(t)\rangle|$ (brown) of the oscillating states.
}
\end{figure*}

We will describe practical ways to realize such systems soon below but first discuss the consequences of the following general observation.
For any solution $\Psi(t)$ of Eq.~\eqref{eq:nonlinwaveeq}, we obtain a partner solution
\begin{equation}
\tilde\Psi(t)=X\Psi^*(t).
\label{eq:partnersolution}\end{equation}
Given our assumptions this correspondence even applies when $H$ and $F$  contain an explicit time dependence
\cite{Note1};  however, we focus on the autonomous case.

Let us first reflect on the possible classes of solutions admitted by Eq.~\eqref{eq:nonlinwaveeq}.
Because of the underlying charge-conjugation symmetry of $H$, the linear system possesses pairs of solutions $\Psi(t)=\Psi_n\exp(-i\Omega_n  t)$ and $\tilde \Psi(t)=X\Psi^*_n\exp(i\Omega^*_n  t)$, where the latter expression corresponds to a frequency $\tilde \Omega_n=-\Omega_n^*$. This includes the possibility of topologically protected zero modes with purely imaginary frequency $\Omega_n=\tilde \Omega_n=i\,\mathrm{Im}\,\Omega_n$ and time dependence  $\Psi(t)=\tilde \Psi(t)=\Psi_n\exp(\mathrm{Im}\,\Omega_n t)$ \cite{PhysRevLett.110.013903,schomerus_topologically_2013,poli_selective_2015,Pikulin2012a,PhysRevB.87.235421,PhysRevLett.115.200402,sanjose2016,sanjose2018}.
If for any of these states $\mathrm{Im}\,\Omega_n$ is positive the linear system is unstable, but the nonlinear effects can stabilize the system.
This results in stationary states $\Psi(t)=\Psi_n\exp(-i\Omega_n t)$ with real $\Omega_n$, which are self-consistent solutions of the condition \cite{Note2}
\begin{equation}
\Omega_n\Psi_n=H\Psi_n+F[\Psi_n,\Psi_n^*]\Psi_n.
\end{equation}
The solutions either

$\quad$(a) occur in pairs $\Psi_n$, $\tilde \Psi_n$ where $\Omega_n$ and $\tilde\Omega_n=-\Omega_n$  now are both real, or

$\quad$(b)  are time-independent zero modes $\Psi_0=\tilde \Psi_0$  whose frequency $\Omega_0=0$ now vanishes
\cite{Note3}.

Alternatively, the system may tend to time-dependent solutions, including periodic, aperiodic and chaotic solutions. These will either

$\quad$(c) still occur in pairs $\Psi(t)$, $\tilde\Psi(t)$ that bear no further relation besides Eq.~\eqref{eq:partnersolution},
or may be constrained in two possible ways:

$\quad$(d) The time-dependent solution may be self-symmetric, $\Psi(t)=\tilde \Psi(t)$; given this condition at some point of time, it will remain true for all times
\cite{Note4}.

$\quad$(e) Two partner solutions may be related by a finite time-offset, $\Psi(t)=\tilde \Psi(t+T/2)$. It then follows that the solutions must be periodic, $\Psi(t+T)=\Psi(t)$, which amounts to the twisted states mentioned in the introduction.

The nature of a state
can be assessed, e.g., by comparing the correlation functions $C=|\langle \Psi(0)|\Psi(t)\rangle|$, $\tilde C=|\langle \Psi(0)|\tilde \Psi(t)\rangle|$, which coincide or alternate for self-symmetric or twisted solutions.

Based on the possible invariance of time-parameterized orbits under discrete involutions, this categorization is minimal and complete.
Note that the solutions of class (a), (d) and (e) all describe orbits $\Psi(t)$ that are invariant under the symmetry operation \eqref{eq:partnersolution}. In the self-symmetric cases (a) and (d), this invariance is local at every point along the orbit, while for the twisted states (e) the invariance occurs under a translation by $T/2$. The latter leaves the orbit invariant as the solution is periodic. All these solutions are furthermore symmetry-protected, in that their number can only change due to structural reconfigurations of the dynamical solutions. That such changes are indeed possible at all is necessary in order to have a setting with meaningful topological features, and not just symmetry-imposed constraints. In linear systems, the phase transitions at which the number and nature of symmetry-protected solutions changes involves degeneracies; as we will explore further in Section \ref{sec:topex}, for the nonlinear setting this naturally extends to bifurcations. First, however, we provide concrete model systems with the time-preserving symmetry of Eq.~\eqref{eq:nonlinwaveeq} and examples of the different classes of solutions.

\begin{figure*}[t]
\includegraphics[width=\textwidth]{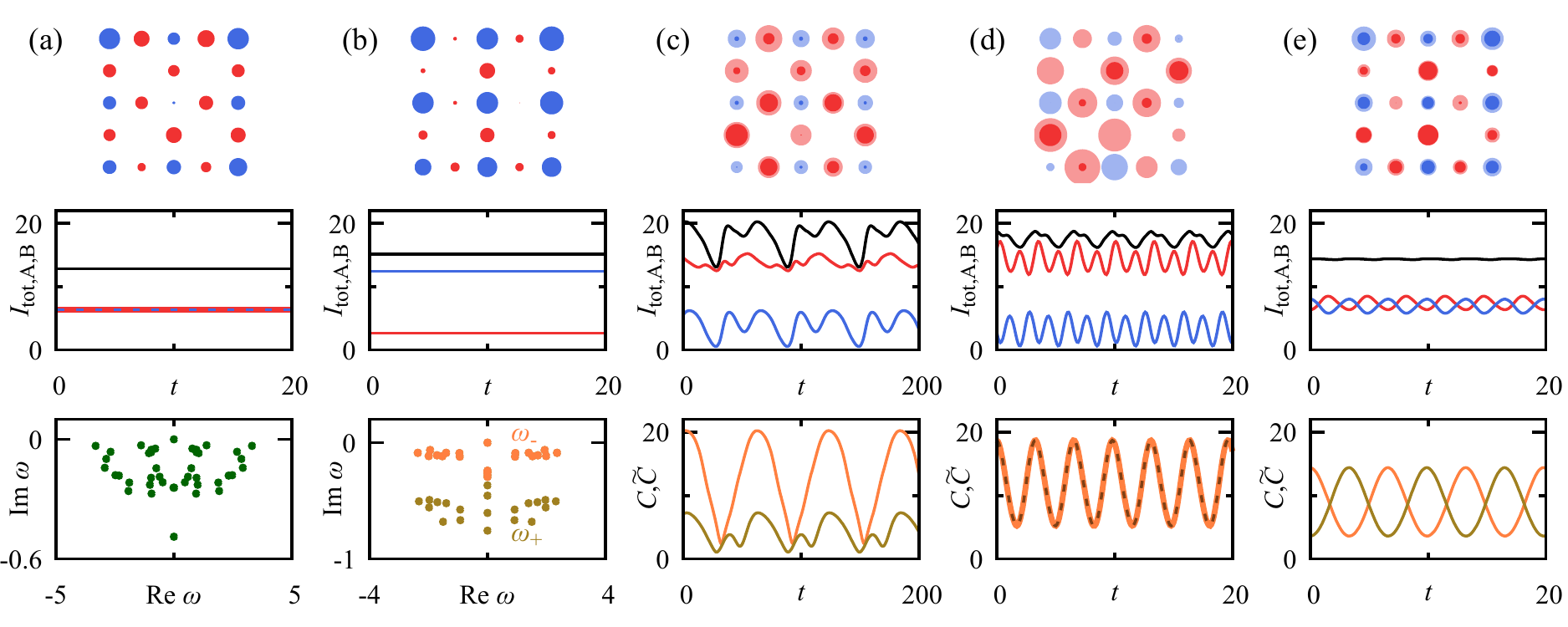}
\caption{\label{fig2}Same as Fig.~\ref{fig1} but
for a polaritonic flat-band condensate based on a Lieb lattice with linear gain and density-dependent loss [see Eq.~\eqref{eq:saturableterms2}]. To demonstrate the generality of our findings we include 50\% relative disorder in all parameters, including for the couplings around their mean value $t_{kl}=1$, and the losses with average strength $\gamma_A=\gamma_B=0.3$ (see Fig.~\ref{fig:app} for details of the configuration).
(a) Stationary symmetry-breaking mode (average gain $g_A=0.15$, $g_B=0.3$). (b) Stationary self-symmetric zero mode (average background loss $g_A=-0.2$ and gain $g_B=0.5$). (c) Slowly oscillating symmetry-breaking mode ($g_A=0.35$, $g_B=0.2$). (d) Oscillating self-symmetric mode ($g_A=0.5$, $g_B=0.3$ with symmetry-preserving initial conditions). (e) Twisted oscillating mode ($g_A=0.1$, $g_B=0.4$).
}

\end{figure*}

\subsection{Realization in lasers and condensates}

\begin{figure}[!ht]
\center{\includegraphics[width=0.8\columnwidth]{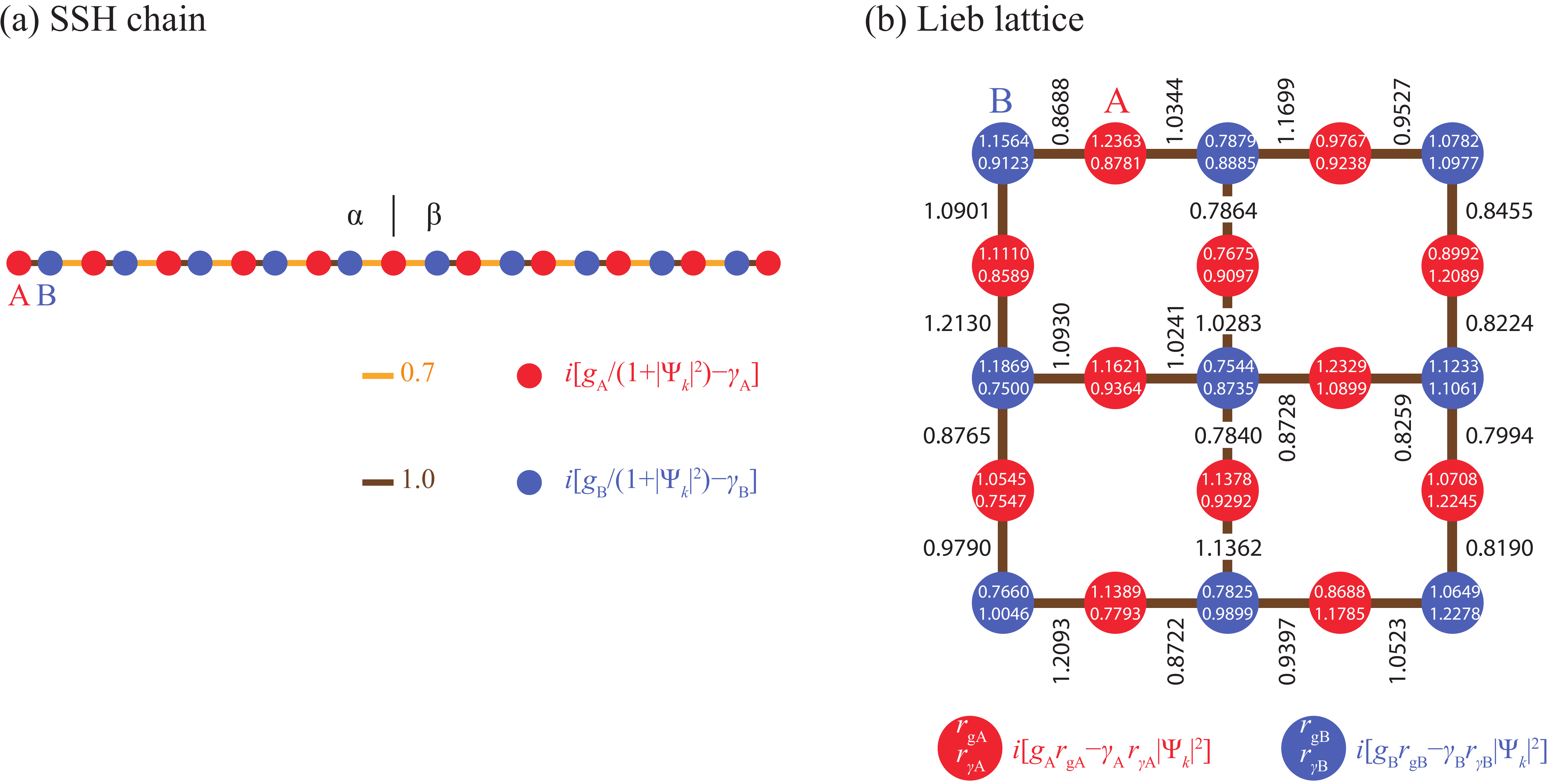}}
\caption{\label{fig:app} Detailed geometry of the two illustrative models investigated in this work. (a) The model based on the SSH chain consists of a linear arrangement of 21 sites (11 A sites and 10 B sites) with alternating couplings 1, 0.7. The centre contains a defect with two consecutive couplings 0.7. This separates two configurations, denoted $\alpha$ and $\beta$, which can be characterized by topological features of their band structure. In the hermitian model the coupling defect induces one zero mode, which is spatially localized and exhibits an anomalous response to loss and gain of different strength on the two sublattices. For this model, we introduce nonlinearities in the form of  saturable gain. (b) The model based on the Lieb lattice also consists of 21 sites, but these are arranged in two dimensions so that 12 are A sites and 9 are B sites. In the hermitian limit, there are now at least three zero modes, even in the case of disorder in the couplings. In this model we study density-dependent losses, and include disorder in the couplings $t_{kl}$ as well as in the gain and loss parameters $g_{A,B}$ and $\gamma_{A,B}$. This disorder is generated from independent random numbers $r_n$ with a box distribution over $[0.75,1.25]$, so that $p_n=p r_n$ for any model parameter $p_n$ with average $p$.}
\end{figure}

Figures \ref{fig1}  and \ref{fig2} illustrate the different types of solutions for two model systems, which are both formed of bipartite lattices ($N_A$ sites A and $N_B$ sites B) that give rise to a pseudospin degree of freedom. The dynamics is governed by the equations
\begin{align}
i \frac{d}{dt} A_k= \sum_{<l>}t_{kl}B_l +f_A(|A_k|^2)A_k,
\label{eq:TB1} \\
i \frac{d}{dt} B_k= \sum_{<l>}t_{lk}A_l +f_B(|B_k|^2)B_k,
\label{eq:TB2}%
\end{align}
where we grouped the amplitudes on both sublattices into vectors $\mathbf{A}$, $\mathbf{B}$.
The real nearest-neighbour couplings $t_{kl}$ form a linear Hamiltonian $H$ that supports
at least $\nu=|N_A-N_B|$ zero modes, irrespective of whether the system is homogenous or not \cite{PhysRevB.34.5208}.
In the examples this describes a finite segment of a Su-Schrieffer-Heeger (SSH) chain with alternating couplings $t_{k,k+1}=1,0.7$
(Fig.~\ref{fig1}), where a topological midgap state arises from a central coupling defect \cite{PhysRevLett.42.1698,ryu_topological_2002,schomerus_topologically_2013}, or a finite segment of a disordered Lieb lattice (Fig.~\ref{fig2}), which exhibits multiple zero modes associated with a flat band \cite{PhysRevLett.62.1201,PhysRevB.54.R17296}.
Both of these models can be realized on a large variety of platforms, including bosonic systems \cite{malkova_observation_2009,poli_selective_2015,%
PhysRevB.81.041410,PhysRevA.82.041402,PhysRevA.83.063601,atala_direct_2013,%
PhysRevLett.114.245503,PhysRevLett.114.245504,PhysRevLett.115.040402,%
PhysRevLett.116.066402,PhysRevLett.116.200402,2053-1583-4-2-025008,2017arXiv170407310S,2017arXiv170503006,R1,R2,R2a}. The detailed geometric configurations in both models are given in Fig.~\ref{fig:app}.

For the SSH chain, we consider nonlinearities that represent saturable gain,
\begin{align}
f_s(|\Psi_k|^2)= i\frac{g_s}{1+|\Psi_k|^2}-i\gamma_s \quad\mbox{(saturable gain)},
\label{eq:saturableterms1}
\end{align}
as often adopted in laser models  \cite{R6},
while for the Lieb lattice we consider density-dependent loss,
\begin{align}
f_s(|\Psi_k|^2)= i(g_s-\gamma_s|\Psi_k|^2) \quad\mbox{(density-dependent loss)}
\label{eq:saturableterms2}
\end{align}
as considered, e.g., in studies of polaritonic condensates  \cite{R7}.

For any solution with amplitudes $\mathbf{A}(t)$ and  $\mathbf{B}(t)$, Eqs.~\eqref{eq:TB1} and \eqref{eq:TB2}, exhibits a partner solution with amplitudes
$\mathbf{A}^*(t)$ and $-\mathbf{B}^*(t)$, so that $X=\mathbb{1}_A\oplus(-\mathbb{1}_B)$ acts as a Pauli-$z$ matrix in pseudospin space.
Admitting different amounts of gain $g_A$, $g_B$ and loss  $\gamma_A$, $\gamma_B$  on the two sublattices allows us to study the competition between topological and nontopological states, which leads to the different examples shown in the figures (for a detailed mapping of the phase space in the SSH model see \cite{Malzard2018}).

Note that the form of $X$ in these models implies that self-symmetric states display a rigid phase-shift of $i$ between the two sublattices. Given that $\tilde \Psi(t)=\Psi(t)$, the amplitudes $\mathbf{A}(t)$ are real while the amplitudes $\mathbf{B}(t)$ are imaginary, which then persists for all times. On the other hand, symmetry-breaking stationary states with a finite frequency must have equal weight on both sublattices  \cite{schomerus_topologically_2013}. The different types of states can therefore also be discriminated by their distinct spatial features.

\subsection{Phase transitions and topological features of the excitation spectrum}
\label{sec:topex}
The presented classification of solutions is exhaustive in terms of symmetry-protection.
To complete the analogy to topological notions in linear systems, it is essential to explore how the number of symmetry-protected solutions can change. In linear systems, this is generally connected to degeneracies, relying on the structural stability of the spectrum against parametric changes. In nonlinear systems, this is paralleled by the notion of structural stability of dynamical solutions (again against parameter changes), with qualitative changes mandated by bifurcations. In this context, we can then exploit the general link of bifurcations to a change of the linear stability (against initial conditions), which allows an emerging spectral interpretation in terms of the stability of linear perturbations.
To conclude this paper,  we therefore further illuminate the topological aspects of the symmetry-protected solutions by spectral signatures. Here, we in particular exploit the pinning of excitations to symmetry protected positions, in analogy to Majorana zero modes in fermionic systems with charge-conjugation symmetry \cite{RevModPhys.82.3045,RevModPhys.83.1057} and analogous zero modes in periodically driven systems \cite{PhysRevB.82.115120,R5}.
These connections can be established by
utilizing the link between dynamical stability of nonlinear systems and their excitation spectrum, which is addressed by Bogoliubov theory. We first present the results of this analysis, and describe the technical details in the following subsection.

The analysis amounts to identifying the eigenmodes $\psi=(u,v)^T$ of linear perturbations $ue^{-i\omega t}+v^*e^{i\omega t}$, and here results in a spectrum of $2(N_A+N_B)$ excitation frequencies $\omega_m$.
For a stable system all of these excitations must obey $\mathrm{Im}\,\omega_m\leq 0$. For non-self-symmetric pairs of stationary modes $\Psi_n$, $\tilde \Psi_n$, the time-preserving symmetry of Eq.~\eqref{eq:nonlinwaveeq} implies that their excitation spectra are identical, but does not impose any further constraints on these spectra.
For self-symmetric stationary zero modes $\Psi_0=\tilde \Psi_0$ with  $\Omega_0=0$, on the other hand, we can classify the perturbations into symmetry-preserving modes $v_+=u_+$ and symmetry-breaking modes $v_-=-u_-$.
For our models of gain or loss, these perturbations fulfill the reduced eigenvalue equations
\begin{align}
&\omega_+ u_+ = (H+f+2f'|\Psi_0|^2)u_+,
\label{eq:evenoddpert1}%
\\
&\omega_- u_- = (H+f)u_-,
\label{eq:evenoddpert2}%
\end{align}
where $f$ is diagonal with entries $f_A(|A_k|^2)$ or $f_B(|B_k|^2)$ (depending on the sublattice), and the prime denotes the derivative with respect to the argument. The excitation spectrum is thus composed of two parts, eigenvalues $\omega_+$ that display an enhanced nonlinearity and eigenvalues $\omega_-$ that coincide with those of the nonlinear wave operator $H+f$, with $\Psi=\Psi_0$ fixed to the zero mode.

Panel (b) in Figs.~\ref{fig1} and \ref{fig2} shows that the eigenvalues $\omega_-$ determine the stability of the stationary zero modes while the eigenvalues $\omega_+$ lie deep in the complex plane. Therefore, if we restrict the perturbations to preserve the symmetry, the stability of such modes is greatly enhanced.
Note that both reduced spectra in these examples have an odd number of excitations. Thus, each reduced spectrum has an odd number of eigenvalues pinned to the imaginary axis. Setting $u_-=\Psi_0$ we recover that one of the eigenvalues $\omega_{-,0}=\Omega_0=0$ vanishes, in accordance with the $\mathrm{U}(1)$ symmetry of the wave equation.

A similar analysis can be carried out for time-dependent solutions $\Psi(t)$, whose stability is then characterized by a corresponding quasienergy spectrum obtained from a propagator $U(t)$ with eigenvalues $\lambda_m=\exp(-i\omega_m t)$.
This again includes the $\mathrm{U}(1)$ Goldstone mode with $\lambda_0=1$, but also an additional Goldstone mode $\lambda_T=1$ that describes the time-translation freedom of any solution $\Psi(t)$.
For general pairs of states $\Psi(t)$, $\tilde\Psi(t)$, the time-preserving symmetry again dictates that their quasienergy excitation spectra are identical.
For self-symmetric states $\Psi(t)=\tilde \Psi(t)$, we can once more separate perturbations  that preserve or break the symmetry. This leads to time-dependent variants of Eqs.~\eqref{eq:evenoddpert1} and \eqref{eq:evenoddpert2},
\begin{align}
&i\frac{d}{dt} u_+(t) = [H+f(t)+2f'(t)|\Psi(t)|^2]u_+(t),
\label{eq:evenoddpertT1}%
\\
&i\frac{d}{dt} u_-(t) = [H+f(t)]u_-(t),
\label{eq:evenoddpertT2}%
\end{align}
where $f$ and $f'$ inherit their time dependence from $\Psi(t)$.
The stability spectrum thus again contains a component $\omega_-$ inherited from the nonlinear wave operator $H+f$, which includes the vanishing eigenvalue $\omega_{-,0}=0$ (with $u_{-,0}=\Psi(t)$) dictated by the $\mathrm{U}(1)$ gauge freedom.
In addition, there is now also an eigenvalue $\omega_{+,0}\equiv\omega_T=0$ arising from the time-translation freedom of any solution $\Psi(t)$, which is associated with $u_{+,0}=d\Psi(t)/dt$.

\begin{figure}[t]
\includegraphics[width=\textwidth]{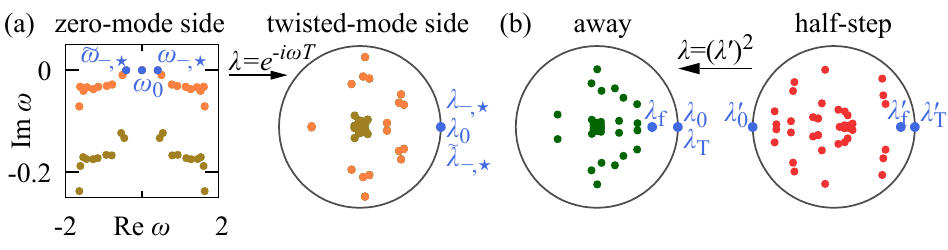}
\caption{\label{fig3} (a) Phase transition from a stationary zero mode to a twisted oscillating state of period $T$ as signalled by the linear stability excitation spectrum,
here shown for the SSH laser array with $g_A=0.693$, $g_B=0.1$.
At the transition the excitations match up via the relation $\lambda=\exp(-i\omega T)$. This leads to a three-fold degeneracy of marginally stable excitations with $\lambda=1$.
(b) Away from the transition ($g_A=0.8$, $g_B=0.1$), the  excitations rearrange to describe the stabilization of the oscillation amplitude of the emerging twisted mode ($\lambda_f$), leaving a two-fold degeneracy of marginal excitations $\lambda_0=\lambda_t=1$ corresponding to $U(1)$ and time-translation invariance. In the half-step operators, these two excitations are separated at $\lambda_0'=-1$, $\lambda_t'=1$ and hence structurally stabilized, which provides a signature of twisted states in terms of topological excitations.
}
\end{figure}

For the final case of a twisted periodic state with $\Psi(t+T/2)=\tilde\Psi(t)$, the evolution operator of the perturbations factorizes as
$U(T)={U'}^2$ with a half-step propagator $U'=\mathcal{X}\Sigma_xU(T/2)$, where $\mathcal{X}=\mathbb{1}_2\otimes X$ while $\Sigma_x=\sigma_x\otimes \mathbb{1}$ interchanges $u$ and $v$.
The $\mathrm{U}(1)$ gauge and time-translation freedoms can now in principle be satisfied in two ways, corresponding to $\omega_{0,T}'T/2=0$ or $\pi$. Inspecting the Goldstone mode
$\psi_0(t)=(\Psi(t),-\Psi^*(t))^T$ we  invariably find that the twisted solutions realize $\omega_0'T/2=\pi$, while the time-translation freedom remains associated with $\omega_T' T/2=0$.
This separation of eigenvalues to symmetry-protected positions is again a robust spectral feature that can only change in further phase transitions.

The key observation in this respect is that these phase transitions naturally link the appearance of twisted modes to the instability of zero modes, which occurs when a pair of excitations
$\tilde\omega_{-,\star}=-\tilde\omega_{-,\star}=2\pi/T$ crosses the real axis  (see Fig.\ \ref{fig3}), and hence corresponds to a Hopf bifurcation \cite{R3,R4} into a time-dependent state that here is still protected by symmetry.
At the transition, the excitations $\omega_n$ map onto the stability
spectrum $\lambda_n=\exp(-i\omega_n T)$  of the emerging twisted state, where $\omega_{-,\star}$ and $\tilde\omega_{-,\star}$ both map onto eigenvalues $\lambda_{-,\star}=\tilde\lambda_{-,\star}=1$. These are degenerate with the $\mathrm{U}(1)$ Goldstone mode $\lambda_0=1$. Away from the transition, these three excitations rearrange into a decaying excitation $|\lambda_f|<1$ describing the stabilization of the oscillation amplitude of the twisted mode, and the Goldstone modes $\lambda_0=\lambda_T=1$ from $\mathrm{U}(1)$ gauge and time-translation invariance.
These phase transitions therefore provide the mechanism by which the proposed class of nonlinear systems acquires robust dynamical features that do not have a counterpart in the linear case.

\subsection{Detailed derivation of the symmetry-constrained excitation spectrum}
We here provide the technical details on the derivation of the stability excitation spectrum for the different solutions identified in the previous subsection.

For stationary states, their stability can be analyzed by setting
\begin{equation}
\Psi(t)=
[\Psi_n+  u \exp (-i\omega t) + v^* \exp (i\omega t)] \exp(-i\Omega_n t)
\label{eq:bgpert}
\end{equation}
and expanding in the small quantities $u$ and $v$, which we group into a vector
$\psi=(u,v)^T$.
This leads to the Bogoliubov equation
\begin{equation}\label{eq:statB}
\omega \psi =(\mathcal{H}[\Psi_n]-\Sigma_z\Omega_n) \psi
\end{equation}
where  $\Sigma_z=\sigma_z\otimes\mathbb{1}$ is a Pauli matrix in the space of $u$ and $v$ while
the Bogoliubov Hamiltonian is given by $\mathcal{H}_{uu}=-\mathcal{H}_{vv}^*=H+(F \Psi)_\Psi$, $\mathcal{H}_{uv}=-\mathcal{H}_{vu}^*=(F \Psi)_{\Psi^*}$.
In our models with saturable gain or density-dependent loss [Eq.~\eqref{eq:saturableterms1} and \eqref{eq:saturableterms2}], this becomes
\begin{align}
&\mathcal{H}[\Psi]=
\left(\begin{array}{cc} H+f+f'|\Psi|^2&f'{\Psi}^2\\ -[f'{\Psi}^2]^* & -[H+f+f'|\Psi|^2]^* \end{array}\right),
\end{align}
where the terms containing $f$ have to be read as diagonal matrix with entries $f_A(|A_k|^2)$ or $f_B(|B_k|^2)$ (depending on the sublattice), and the prime denotes the derivative with respect to the argument.
As any Bogoliubov equation, Eq.~\eqref{eq:statB} exhibits the charge-conjugation symmetry $(\mathcal{H}[\Psi_n]-\Sigma_z\Omega_n)^*=-\Sigma_x(\mathcal{H}[\Psi_n]-\Sigma_z\Omega_n)\Sigma_x$, with the corresponding Pauli matrix $\Sigma_x=\sigma_x\otimes\mathbb{1}$. This implies that the excitation spectrum is symmetric under the transformation $\omega_m\to\tilde \omega_m=-\omega_m^*$.
Furthermore, due to the $\mathrm{U}(1)$ gauge freedom, Eq.~\eqref{eq:statB} always admits the Goldstone mode $\psi_0=(\Psi,-\Psi^*)^T$ with eigenvalue pinned at $\omega_0=0$. As the dimension of $\mathcal{H}$ is even, an additional odd number of eigenvalues must lie on the imaginary axis. We need to explore the interplay of these spectral features with the analogous symmetries in $\Omega$.

The time-preserving symmetry of the nonlinear wave equation \eqref{eq:nonlinwaveeq} induces the relation
\begin{equation}
\mathcal{X}(\mathcal{H}[\Psi_n]-\Sigma_z\Omega_n)^*\mathcal{X}=-(\mathcal{H}[\tilde\Psi_n]+\Sigma_z\Omega_n)
\label{eq:bgsym}
\end{equation}
where $\mathcal{X}=\mathbb{1}_2\otimes X$. For non-self-symmetric pairs of stationary modes $\Psi_n$, $\tilde \Psi_n$, this implies that their excitation spectra are identical.
For self-symmetric stationary zero modes $\Psi_0=\tilde \Psi_0$ with  $\Omega_0=0$, on the other hand, Eq.~\eqref{eq:bgsym} turns into an additional charge-conjugation symmetry which imposes a constraint on the excitation spectrum. In this case, we can classify the perturbations into joint eigenstates of $\mathcal{X}$ with eigenvalue $\sigma=\pm1$. These take the form $v_\sigma=\sigma X u_\sigma$, and fulfill the reduced eigenvalue equations
\begin{equation}
\omega_\pm u_\pm=\left[H+F+F_\Psi\Psi_0\pm F_{\Psi^*}\Psi_0X\right]u_\pm.
\end{equation}

For our models with saturable gain or density-dependent loss, this simplifies to
Eqs.~\eqref{eq:evenoddpert1} and \eqref{eq:evenoddpert2}. As described there,
the excitation spectrum is thus composed of two parts, eigenvalues $\omega_+$ that display an enhanced nonlinearity and eigenvalues $\omega_-$ that coincide with those of the nonlinear wave operator $H+f$, with $\Psi=\Psi_0$ fixed to the zero mode. According to Eq.~\eqref{eq:bgpert}, the perturbations $[e^{-i\omega_- t}u_--e^{i\omega_- t} Xu_-^* ]$ corresponding to $\omega_-$ break the time-preserving symmetry of the zero mode.
Setting $u_-=\Psi_0$ we recover that one of the eigenvalues $\omega_-=\Omega_0=0$ vanishes, in accordance with the $\mathrm{U}(1)$ symmetry of the wave equation.

For time-dependent solutions $\Psi(t)$, their stability against perturbations $u(t) + v^*(t)$ is governed by the time-dependent Bogoliubov equation
 \begin{equation}
i\frac{d}{dt} \psi(t)
=\mathcal{H}[\Psi(t)]\psi(t),
\label{eq:timedepB}
\end{equation}
whose solutions $\psi(t)=U(t)\psi(0)$ define the propagator $U(t)$.
The eigenvalues $\lambda_m=\exp(-i\omega_m t)$ of $U(t)$ can be represented by complex quasienergies that are constrained
in similar ways as the excitations in the stationary case. The conventional charge-conjugation symmetry $\mathcal{H}^*(t)=-\Sigma_x\mathcal{H}\Sigma_x$ implies $U^*=\Sigma_xU\Sigma_x$, so that
each $\omega_m$ comes with a partner $\tilde\omega_m=-\omega_m^*$ (thus $\tilde \lambda_m=\lambda_m^*$) or is purely imaginary (whereupon $\lambda_m$ is real). This includes the $\mathrm{U}(1)$ Goldstone mode $\psi_0(t)=(\Psi(t),-\Psi^*(t))^T$, as well as an additional Goldstone mode $\psi_T(t)=(d\Psi/dt,d\Psi^*/dt)^T$ which describes the time-translation freedom $t_0$ of any solution $\Psi(t+t_0)$.

For general pairs of states $\Psi(t)$, $\tilde\Psi(t)$, the propagators are related as $U(t)=\mathcal{X}\tilde U^*(t)\mathcal{X}=\mathcal{X}\Sigma_x \tilde U(t)\Sigma_x \mathcal{X}$, which again implies that their quasienergy excitation spectra are identical.
This also applies to the case of  periodic solutions with $\Psi(t)=\Psi(t+T)$, where the propagator $U(T)$ represents a Bogoliubov-Floquet operator. Such solutions are unstable if $|\lambda_m|>1$, hence $\mathrm{Im}\,\omega_m>0$, apart from the two Goldstone modes that are fixed at $\lambda_0=\lambda_T=1$.
For the more general case of periodic solutions with $\Psi(t)=\exp(i\varphi) \Psi(t+T)$, this applies when the  Floquet operator is redefined as $\mathrm{diag}(e^{-i\varphi},e^{i\varphi})U(T)$.

For self-symmetric states $\Psi(t)=\tilde \Psi(t)$, we have at all times $U(t)=\mathcal{X}U^*(t)\mathcal{X}=\mathcal{X}\Sigma_x U(t)\Sigma_x \mathcal{X}$. This allows us to separate perturbations  that preserve or break the symmetry, and leads to the time-dependent variants \eqref{eq:evenoddpertT1} and \eqref{eq:evenoddpertT2} of Eqs.~\eqref{eq:evenoddpert1} and \eqref{eq:evenoddpert2}.
For periodic states $\Psi(t)=\exp(i\varphi) \Psi(t+T)$, the self-symmetry constraints the phase to $\varphi=0,\pi$, hence $\tau=\exp(i\varphi)=\pm 1$, which needs to be respected in the proper definition of the Bogoliubov-Floquet operator  $F=\tau U(T)$
to result in two Goldstone modes pinned at $\lambda_0=\lambda_T=1$.

For the final case of a twisted periodic state with $\Psi(t+T/2)=\tilde\Psi(t)$, we have automatically $\Psi(T)=\Psi(0)$ without any additional phase [furthermore, the case $\Psi(t+T/2)=-\tilde\Psi(t)$ is  not separate since we can then redefine $\Psi(t)\to i \Psi(t)$]. The Floquet operator factorizes as
\begin{align}
U(T)=\mathcal{X}U^*(T/2)\mathcal{X}U(T/2)
=\mathcal{X}\Sigma_xU(T/2)\Sigma_x\mathcal{X}U(T/2).
\end{align}
In this case, the stability spectrum $\lambda_m=(\lambda_m')^2$ thus follows from the eigenvalues $\lambda_m'$ of the half-step propagator $U'=\mathcal{X}\Sigma_xU(T/2)$. The $\mathrm{U}(1)$ gauge freedom can now in principle be satisfied in two ways, corresponding to $\lambda_0'=\pm 1$. However, inspecting the Goldstone mode $\psi_0(t)=(\Psi(t),-\Psi^*(t))^T$ we invariably find $\psi_0(T/2)=-\mathcal{X}\Sigma_x\psi_0(0)$ so that the twisted solutions indeed always realize the case $\lambda_0'=-1$. The Goldstone mode $\psi_T(t)=(d\Psi/dt,d\Psi^*/dt)^T$ of the time translation, on the other hand, fulfills $\psi_T(T/2)=\mathcal{X}\Sigma_x\psi_T(0)$, and hence always lies at $\lambda_T'=1$.
Note that this spectral separation persists if we were to redefine the half-step propagator as $-\mathcal{X}\Sigma_xU(T/2)$ (as we are entitled to do; both conventions make sense), upon which the two symmetry-protected excitations swap their positions.

\section{Conclusion}
In summary, we showed that spectral symmetries underpinning topological quantum systems can be extended to nonlinear complex-wave equations, where they lead to robust constraints of the dynamics.
Conceptually, this allowed us to identify a generalization of zero modes from the linear to the nonlinear setting, as well as
 symmetry-protected power-oscillating modes that do not have a counterpart in the linear case. These states furthermore support symmetry-protected excitations, which play a crucial role in their dynamical features.
Practically, nonlinearities are of fundamental importance to stabilize systems with loss and gain in quasistationary operating regimes. In particular,
the symmetry-protected modes can be induced into
weakly interacting bosonic systems with saturable gain or density-dependent loss, such as recently realized in polaritonic condensates \cite{PhysRevX.5.011034,2017arXiv170503006} and topological lasers \cite{R1,R2,R2a}. The results can also be reinterpreted in  classical wave settings, as encountered, e.g., in conventional optics and acoustics.

These findings show that the notion of symmetry-protected topological states persists in nonlinear bosonic systems, and that the nonlinearities lead to an even richer phenomenology than in the original single-particle context.
On the practical side, it is worthwhile to explore further extensions to include
the dynamics of the medium \cite{Longhi2017}, which is particularly important when one considers time-dependent solutions. More generally, it would be highly worthwhile to explore nonlinear extensions for other representatives of linear topological universality classes, for which experimental realizations and a general classification are currently unknown.

\section*{Funding}
UK Engineering and Physical Sciences Research Council (EPSRC) (EP/N031776/1, EP/P010180/1).
The research data is accessible at doi 10.17635/lancaster/researchdata/235


\section*{Disclosures}
 The authors declare that there are no conflicts of interest related to this article.

\end{document}